\documentclass[a4paper,fleqn,usenatbib,useAMS]{mnras}

\usepackage{graphicx}	
\usepackage{amsmath}	
\usepackage{amssymb}	
\usepackage{multicol}        
\usepackage{bm}		
\usepackage{pdflscape}	

\usepackage[T1]{fontenc}
\usepackage{ae,aecompl}

\usepackage{txfonts}

\title[A85, A2457, IIZw108 an IFU view]{Stellar Populations of BCGs, Close Companions and Intracluster Light in Abell~85, Abell~2457 and IIZw108}

\author [Edwards et al.]{L. O. V. Edwards$^1$, H. S. Alpert$^1$, I. L. Trierweiler$^1$, T. Abraham$^1$, V. G. Beizer$^1$\\
$^1$Department of Astronomy, Yale University, New Haven, CT\\
}
\date{\today}
\begin{document}
\maketitle
\begin{abstract}

We present the first results from an integral field (IFU) spectroscopic survey of a $\sim$75$\,$kpc region around three Brightest Cluster Galaxies (BCGs), combining over 100 IFU fibres to study the intracluster light (ICL).  We fit population synthesis models to estimate age and metallicity. For Abell~85 and Abell~2457, the ICL is best-fit with a fraction of old, metal-rich stars like in the BCG, but requires 30-50\% young and metal-poor stars, a component not found in the BCGs. This is consistent with the ICL having been formed by a combination of interactions with less massive, younger, more metal-poor cluster members in addition to stars that form the BCG. We find that the three galaxies are in different stages of evolution and may be the result of different formation mechanisms. The BCG in Abell~85 is near a relatively young, metal-poor galaxy, but the dynamical friction timescale is long and the two are unlikely to be undergoing a merger. The outer regions of Abell~2457 show a higher relative fraction of metal-poor stars, and we find one companion, with a higher fraction of young, metal-poor stars than the BCG, which is likely to merge within a gigayear. Several luminous red galaxies are found at the centre of the cluster IIZw108, with short merger timescales, suggesting the system is about to embark on a series of major mergers to build up a dominant BCG. The young, metal-poor component found in the ICL is not found in the merging galaxies.

\end{abstract}
\begin{keywords}
galaxies: evolution; galaxies: interactions; galaxies:cD; galaxies: clusters.
\end{keywords}

\section{Introduction}
Brightest cluster galaxies (BCGs) are the most massive and luminous galaxies in the local Universe. They are typically cD galaxies found at the centre of galaxy clusters and are dominated by old, low-mass stars. The hierarchical build-up model of BCGs is supported both by cosmological simulations \citep[]{dub98,del07,ton12} and observations of multiple nuclei and many close pairs among cD galaxies \citep[]{ost77}. An understanding of the timescale and mechanisms behind this process may be obtained by investigating the cluster environment, as well as the stellar populations and dynamics of the BCG, its close companions and the intracluster light (ICL).

An observational study of 433 low-redshift BCGs \citep{lau14} is consistent with the cosmological simulations of \citet{mar14} in that most of the stellar mass of BCGs was not assembled inside the rich cluster environment. Rather, many progenitor galaxies formed within infalling groups and then merged to form a larger galaxy, which was later accreted by the cluster.  The idea of downsizing \citep[]{cow96,smi06}, also plays a role in the formation of today's largest galaxies - these would have formed their stars early on in the dense regions of the Universe \citep[]{tof14,nel14}. If most of the stars in the BCGs formed long ago, we would expect the BCG core to be old and have high metallicity \citep{nel05,bro07,van15}. Evidence for this has been gathered by \citet{lou09} and \citet{von06}, who find that the cores of BCGs are dominated by old, metal-rich populations. However, the picture may not be so simple. BCG cores have been found to be host to a range of metallicities \citep{bro07}, and many in \citet[]{gro14}'s sample of 39 BCGs required complex star formation histories. Many have emission lines \citep{cra99,hat07,edw09,mcd12}. X-ray selected clusters are more likely to show star formation or AGN activity for their BCGs \citet{liu12}, and much of this star formation activity has been associated with cool core systems \citep[]{bil08,edw07,gro14}.

The preponderance of nearby neighbours may also be correlated with recent BCG growth. In optically selected systems, a dominant galaxy which is starforming tends to be in a group (BGGs, \citet{oli14}). But, dominant cluster galaxies have also shown evidence for continuing growth.  For example, minor mergers today may contribute to the BCG's mass \citep{mci08,edw12,ina15, lon15}.  Furthermore, a recent study by \citet{oli15} has shown a lack of metallicity gradients in local BCGs, more evidence for recent merging.  \citet[]{bro11} show that BCG major mergers are possible at low redshift, and \citet[]{liu12} find merger features in a sample of nearby optically selected starforming BCGs. This indicates that although most of the BCG formation has occurred in the distant past, there should still still clusters with BCGs currently undergoing merging activity.

Another important component of the cluster core closely connected to the evolutionary history of the system is the intracluster light (ICL). The ICL is a diffuse collection of stars that are thought to be bound to the gravitational well of the cluster, but overlap with the location of the BCG. Estimates of the amount of ICL that exists in clusters varies, with studies finding anywhere from 5$\%$ to 50$\%$ \citep{mcg10,dar08,zib05,bur15} of the light in galaxy clusters belonging to the ICL.  Whereas most of the mass of the BCG is in place by z=1, the ICL has grown by a factor of 4-5 since z$<$0.4 \citep{bur15}. It is unknown whether the formation of BCGs is concurrent with the formation of the diffuse ICL \citep{fel04,bur12}, but large photometric surveys suggest the ICL evolution is connected more to the cluster as a whole than to the BCG \citep[]{lin04,gon05}, formed from interactions with cluster galaxies, through mergers of galaxies in the core, or infalling systems. The models and simulations of \citet{con07}, \citet{puc10} and \citet{lap13} all predict that the bulk of the ICL is formed through tidal stripping of infalling massive galaxies, and \citet{con14} find that a component from merging cluster galaxies can also be important when included (see also \citet{mur07}). These predict sub-solar metallicities, lower than what is seen in BCGs \citep{von06}. One plausible component of the ICL is an addition from mergers that built the BCG, though there are not enough to account for the total amount of ICL in clusters \citet{bur15}. The ICL therefore may have components from both processes: from tidal stripping of infalling galaxies (which might be associated with cluster buildup) and where we would expect a metal-poor component from the field galaxies, as well as from merging events with the BCG (which could be associated with BCG formation) and where we would expect the stellar populations to reflect the outer BCG envelope.  Indeed, from deep Hubble Space Telescope imaging, \citet[]{mon14} recently showed that the stars of the ICL in Abell~2744 are much younger than those of the BCG and other massive cluster galaxies, which is different to the spectroscopic observations of \citet{mel12}. This is currently unresolved as the spectroscopic observations of the ICL currently available from studies of clusters do not agree. \citet{mel12} and \citet{tol11} find that the ICL spectrum of a z$\sim$0.3 cluster does not fit with a population dominated by old, metal-rich stars, and that young, metal-poor stars can not make up more than 1\% by mass. This is in stark contrast to the spectroscopic results of \citet{coc11} who looked at the nearby Hydra~I system finding the population to be dominated by old, metal-poor stars, and to the work on Coma \citep{coc10}, which finds the stars of the ICL to be old with solar metallicity. By examining the stellar populations of the ICL in conjunction with the BCG and close companions, for a sample of clusters, we can test the formation scenarios. What is needed is a sample of spectroscopically studied ICL studied consistently, along with the BCG and potential mergers.

We are building such a sample from collecting IFU data on nearby cluster cores. Here we present results for three local systems: Abell~85, Abell~2457, and IIZw108. This is the first time the ICL stellar populations have been derived from IFU data.

In Section~2, we discuss our sample selection, observations, and data reduction and discuss how the population synthesis code STARLIGHT \citep{cid05} is used to fit a suite of model spectra from \citet{wal09} to our observations. In Section~3, we report our results: age and metallicity from the best-fit model, across the BCG, for the companions and in the ICL. We also estimate the merger timescale and likelihood. In Section~4, we compare our results to those in the literature and discuss the results in terms of BCG and ICL formation scenarios. In Section~5 we conclude.  H$_{o}$=67.8 km/s/Mpc, $\Omega_{m}$=0.3, and $\Omega_{\Lambda}$=0.7 throughout.

\begin{table*}
\caption{Observational Data. Each target is observed in 3 positions: Home (H), and two offsets (O1, O2) from the home position. }
\label{tab:ObservationalData}
\begin{tabular}{l l l l c c c c c}
\hline \hline
Cluster & Dates & RA & Dec & IT & Rot. Pos. & FOV & E(B-V) & Standard\\
& & (h:m:s) & ($^{o}$:$^{\prime}$:$^{\prime\prime}$) &  (s) & (deg) & (kpc$^2$) & &Star\\ \hline
A85 H & 11-25-13 & 00:41:50.54 & -09:18:13.07  & 3600 & -15 & 77.5$\times$76.8 & 0.034 &PG0310149\\
A85 O1 & 11-25-13 & H-1.46$^{\prime\prime}$ & H+5.44$^{\prime\prime}$  & 3600 & -15 & &&PG0310149\\
A85 O2 & 11-25-13 & H-5.50$^{\prime\prime}$ & H+1.44$^{\prime\prime}$ & 3600 & -15& &&PG0310149\\
A2457 H & 10-13-14 & 22:35:40.74 & +01:29:05.76  & 3600 & 0 & 83.2$\times$82.4 & 0.074 &HD217086\\
A2457 O1 & 10-13-14 & H+0$^{\prime\prime}$ & H-5.63$^{\prime\prime}$ &  3600 & 0 &&& HD217086\\
A2457 O2 & 10-13-14 & H+4.90$^{\prime\prime}$ & H-2.80$^{\prime\prime}$  & 3600 & 0& &&HD217086\\
IIZw108 H & 10-15-14 & 21:13:54.79 & +02:33:51.00 & 3600 & 103&70.0$\times$69.3 &0.060 & HD217086\\
IIZw108 H & 11-27-13 & 21:13:55.21 & +02:33:52.00 & 5400 & 102&  & &PG0310149 \\
IIZw108 O1 & 11-28-13 & H-5.89$^{\prime\prime}$ & H-1.27 $^{\prime\prime}$ &  3600 & 102&&&PG0310149\\
IIZw108 O2 & 11-28-13 & H-3.85$^{\prime\prime}$ & H+4.17$^{\prime\prime}$ & 3600 & 102&&&PG0310149\\
\hline
\end{tabular}
\end{table*}

\section{Observations and Data Reduction}

\subsection{Sample Selection}
We use BCGs drawn from the NOAO Fundamental Plane (NFP) survey \citep[]{smi04}, which examined 93 of the brightest X-ray clusters in the ROSAT all sky survey \citep[]{ebe98}. From the 61 northern clusters, 25 cool-core and non-cool-core clusters that lie in the redshift slice z$=$0.045 to 0.07 were selected, thus allowing a field of view of 30-100$\,$kpc around the BCG to be observed in one pointing. This region is occupied by close companions that are likely to merge within a gigayear \citep{edw12} and stretches to where a contribution from intracluster light is expected \citep{pre14}. Sixteen clusters have thus far been observed, and in this first paper, we present three representative targets in order to discuss the techniques used and highlight the potential of the survey.  

Table~\ref{tab:ObservationalData} lists these clusters which represent the diversity that exists in the larger sample. Abell~85 is a massive, relaxed, cool-core cluster \citep[]{per98} with no relatively bright close companions. Abell~2457, similarly, has no bright companions, and has evidence for only a weak cool-core \citep[]{lak14}. Some substructure has been identified in the cluster \citet[]{fli06}. IIZw108 is an unrelaxed \citep[]{che07}, poor cluster with many bright galaxies in proximity. Table~\ref{tab:ClusterData} lists properties of the host clusters, including whether the cluster has a cool core as well as the difference in r$^{\prime}$-band magnitude of the BCG and next brightest companion galaxy, used as a proxy for the merging status. The magnitudes are from the Sloan Digital Sky Survey (SDSS) fibre observations, thus are underestimates of the total BCG light.

\begin{table}
\caption{Properties of Observed Targets. The cluster name is shown in column 1. Column 2 lists the cooling flow status, column 3 lists the differential magnitudes of 1st and 2nd rank members within the field of view. Column 4 gives the cluster mass and cluster velocity dispersion from \citet[]{smi04}. Abell~2457 does not have enough cluster members with measured redshifts to derive a velocity dispersion. The final column lists the redshift.}
\label{tab:ClusterData}
\begin{tabular}{l l l c c c c}
\hline \hline
Cluster & CF Status & $\Delta$M$_{1}$-M$_{2}$ & L$_{X}$ & $\sigma_{cluster}$ & z\\
&&&10$^{37}$W&km s$^{-1}$&\\
A85&strong&-0.72&4.92&736&0.0551\\
A2457&weak&-2.88&0.45&-&0.0594\\
IIZw108&non-CF&+3.37&1.09&399&0.0493\\
\hline
\end{tabular}
\end{table}

\subsection{Observations}

The BCGs and associated standard stars of Table~~\ref{tab:ObservationalData} were observed with SparsePak \citep[]{ber04} on the 3.5$\,$m WIYN telescope over the course of two observing runs in November 2013 and October 2014.  SparsePak is a sparsely packed quasi-integral field unit with 82 4.7$^{\prime\prime}$ fibres of which 75 are arranged in a grid of dimensions 72$^{\prime\prime}\times$71.3$^{\prime\prime}$. The centre is tightly packed and 7 sky fibres are located on the outside of the grid.  The GG-420 filter was used along with the grating 600@10.1, with angle 25.2$^\circ$, to obtain spectra with a central wavelength of 5703$\,$\AA. Here, the spectral resolution is 3.2$\AA$ per pixel and the dispersion is 1.0$\AA$ per pixel. In November 2013, the spectral range covered was $\sim$4320-7150\AA; for the October 2014 run, the range was $\sim$4270-7100\AA.  The nights varied in their cloud cover and the seeing ranged from about 0.5$^{\prime\prime}$ to 1.5$^{\prime\prime}$, well within the SparsePak fibre diameter.

\begin{figure*}
\begin{minipage}[c]{0.34\textwidth}
\centering
\includegraphics[height=2.4in, angle=0]{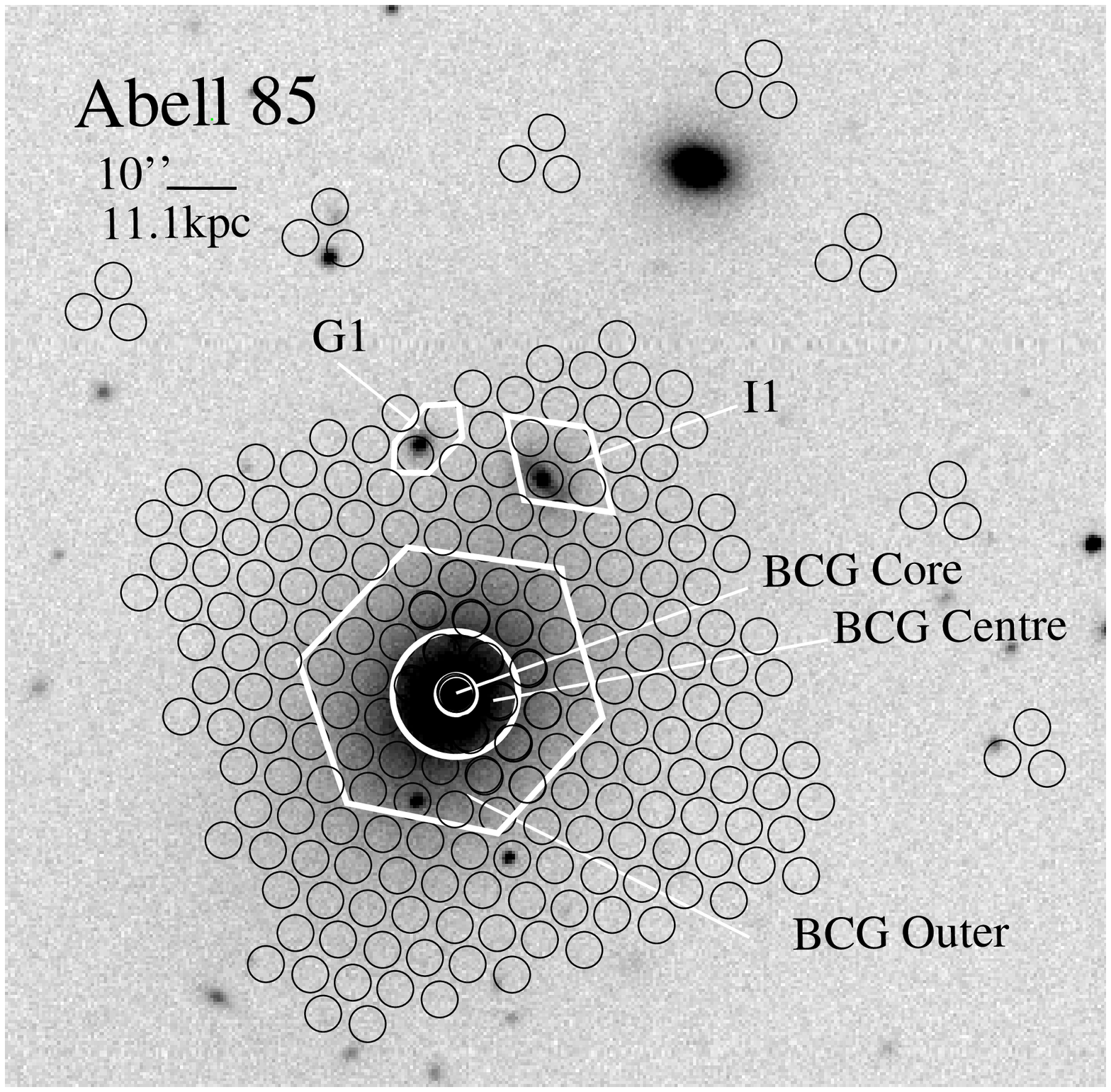}
\end{minipage}%
\begin{minipage}[c]{0.33\textwidth}
\centering
\includegraphics[height=2.4in, angle=0]{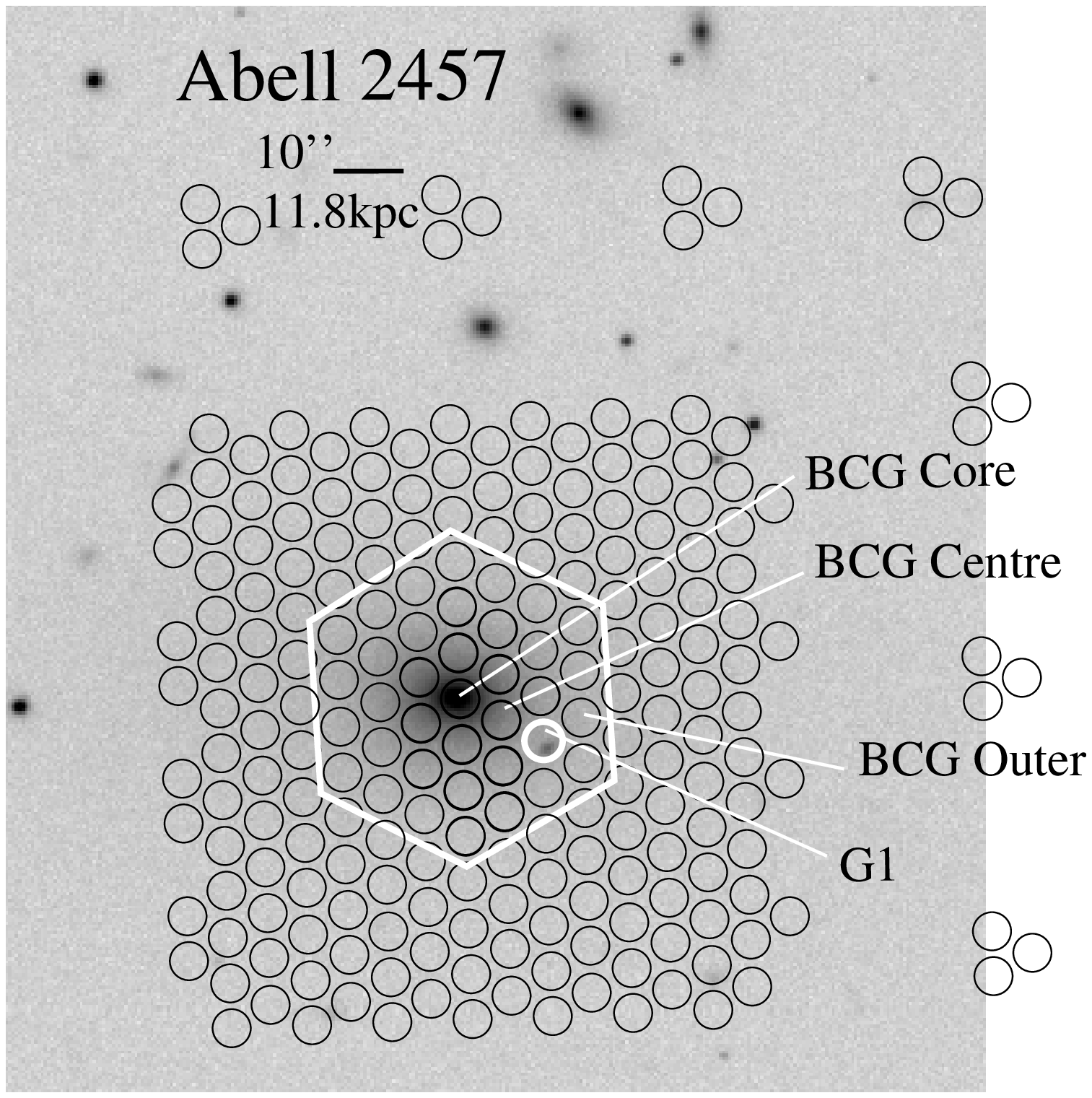}
\end{minipage}%
\begin{minipage}[c]{0.28\textwidth}
\centering
\includegraphics[height=2.4in, angle=0]{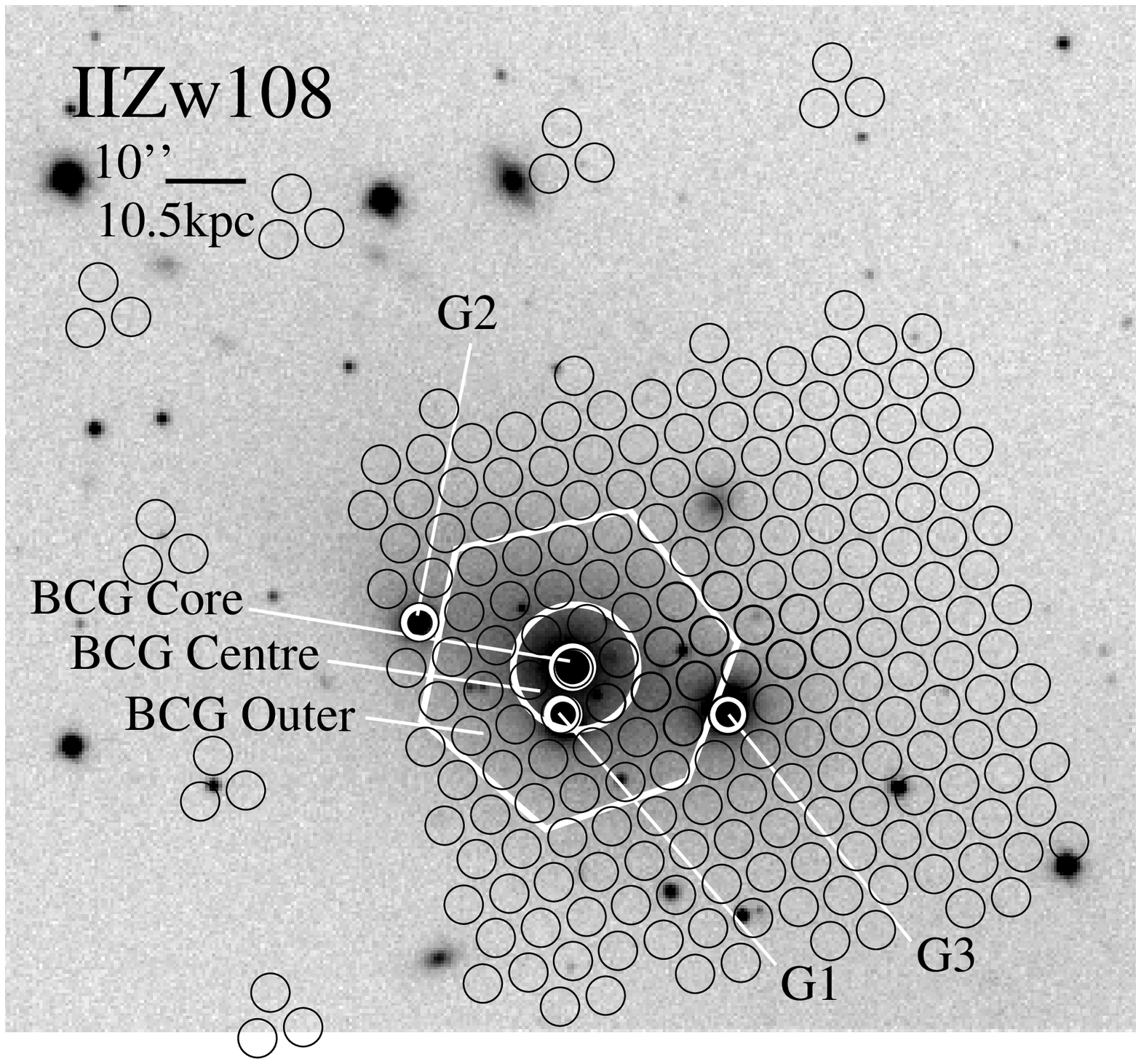}
\end{minipage}%
\caption {\textit{The field of view covered by SparsePak is illustrated on top of greyscale SDSS r$^{\prime}$ images. White polygons illustrate the fibres combined together to construct the region spectra discussed in the text.  Fibres used for sky subtraction are outside of the main bundle.  For Abell~85 (left), the FOV covers 77.5$\times$76.8 kpc$^2$, for Abell~2457 (centre) it covers 83.2$\times$82.4 kpc$^2$ and for IIZw108 (right) it covers 70.0$\times$69.3 kpc$^2$. Interloping galaxies are delineated with an I, and spectroscopically confirmed cluster companion galaxies are labeled with a G. North is up with East to the left.}}
\label{fig:contmaps}
\end{figure*}

Table \ref{tab:ObservationalData} lists the instrument rotation angle, field of view (FOV) and galactic extinction for each target. Three positions, Home (H), offset 1 (O1), and offset 2 (O2) were observed for each target to fully integrate the field and oversample the BCG. At zero declination and a rotator angle of zero, the first offset is 0$^{\prime\prime}$E 5.6$^{\prime\prime}$S and the second is 4.9$^{\prime\prime}$W 2.8$^{\prime\prime}$N from the original position. Offsets specific to our setup are given in the table for each target.  Optical r$^{\prime}$ images from the SDSS are shown in Figure~\ref{fig:contmaps}, illustrating the position and extent of the SparsePak footprint for each cluster. The FOV covered is equivalent to $\sim$70-80$\,$kpc centred on the BCG.

Based on their r$^{\prime}$-band magnitude, sources were integrated so that the 20:1 companions to each BCG could be detected. Observations were performed in blocks of 600s exposures at each position six times to ensure the brightest galaxies were not saturated, and to remove cosmic rays. Individual fibres centred on the BCG have a continuum signal to noise ratio of $\sim$20 per angstrom, which falls to $\sim$2 on the ICL. IIZw108 Home has a slightly higher ratio ($\sim$30 on the BCG core) as it was observed on two different nights, one of which was 9 x 600s.

\subsection{Data Reduction}

The data was reduced using the Image Reduction and Analysis Facility (IRAF) V2.14, specifically the \textit{dohydra} package.\footnote{A guide for the basic data reduction process can be found here: http://www.astro.wisc.edu/$\sim$cigan/reducing/reducing.html} The images were trimmed, bias and dark subtracted. Flat fields for each night were median combined. 

The program L.A. Cosmic \citep[]{van01} was used to remove the majority of the cosmic rays present throughout the data by setting \textit{sigclip=9}, \textit{objlim=10}, and \textit{niter=2}. The \textit{dohydra} package was run over the target files to apply the flat field correction and wavelength calibration. The latter uses comparison spectra from the CuAr lamp, taken each night of the run.

\subsubsection{Sky Subtraction}

We experimented with several sky-subtraction procedures. The first method was subtracting sky fields - of the same integration time and reduced following the same procedure as the targets - from spectra of our objects. This expensive method produced a number of flagrant over- and under-subtractions. Next, the sky spectra were scaled by flux and subtracted from the object spectra. This procedure had little effect in improving the data with many sky lines remaining. The best results relied on using the seven apertures ($\times$3 pointings) of SparsePak that fall outside the main pack of fibres (see Figure~\ref{fig:contmaps}). After excluding those that fell on a foreground or background object, sky fibres for each cluster at every position were median combined. We then subtracted this median sky spectrum from the object spectra. The downside to this method is that if the ICL extends to these fibres, it is effectively subtracted out of the data. However, for the most part, the ICL is too faint to be above the level of the noise in the data that far from the BCG centre. The data was reduced from each of the three nights separately, overlapping positions were median combined, resulting in one sky-subtracted file for each of the home and offset positions for every cluster. Once finished, the remaining cosmic rays were manually removed using the editing tools in \textit{splot}.

\subsubsection{Reddening, Extinction and Flux Calibration}

The IRAF task \textit{deredden} was used to account for extinction from Galactic dust (Table~\ref{tab:ObservationalData}), with no internal dust extinction corrections applied. The package \textit{dopcor} was used to account for the cluster redshift (Table~\ref{tab:ClusterData}). E(B-V) values and redshifts were obtained from the NASA Extragalactic Database (NED).

Both flux-calibrated and normalized spectra were produced. Standard stars were observed each night (Table~\ref{tab:ObservationalData}), reduced and sky-subtracted as above.  IRAF's \textit{calib} task was used to perform the flux calibration. For most of the sample the flux calibration is stable to 4\%, based on measurements of a cluster in the survey that was observed on two different nights without the issue of positional offsets (Abell~407). Observations of IIZw108 in the home position were taken in November and October, with a slight positional and rotational offset between the two. Comparing the flux ratio  at 6700$\,$\AA~(2.46$\times$10$^{-16}$$\,$erg/s/cm$^{2}$/\AA, 2.64$\times$10$^{-16}$$\,$erg/s/cm$^{2}$/\AA) on a fibre which covers the same object, the absolute run-to-run flux calibration is stable to 7\%. The flux-calibrated spectra are then normalized by fitting a high-order polynomial to the continuum, which was divided through each spectrum. 

\subsection{Population Synthesis Models}\label{sec:starlight}

To determine stellar populations, the flux-calibrated spectra are resampled to a 1$\,$\AA~resolution and fit with the STARLIGHT \citep{cid05,cid07,mat06,asa07} population synthesis code. Many authors have found BCGs to be well-fit by $\alpha$-enhanced models \citep{smi04,von06,lou09}.  Therefore, our spectra are fit to the \citet{wal09} models which include three possible $\alpha$-enhancement values (0.0, 0.2 and 0.4). We also explored fits to the $\alpha$-enhaneced \citet{coe07} models and the non-$\alpha$ enhanced \citet{bru03} models, which include younger ages. Although the specific values of age and metallicity changed, the overall trends did not.   \citet{smi04} found that local large red galaxies were older if the galaxies were more massive, thus, we explore 54 single stellar population (SSP) models which cover ages from 3$\times$10$^{9}$ Gyr to 13$\times$10$^{9}$ Gyr and metallicities ranging from 0.006, 0.020 and 0.032. The top-heavy \citet{cha03} initial mass function is used as well as the \citet{cal00} reddening law. 

STARLIGHT then runs a suite of monte-carlo simulations varying the amount of each base spectrum to be mixed into the model fit, from which the best-fit model is determined by comparing to the observed spectrum from 4100\AA-6750\AA ~and choosing the model with the lowest chi-squared value. The program also measures the velocity dispersion. Our input spectra are corrected for instrumental broadenning, and deredshifted to the cluster velocity, so the shifts calculated refer to the offset with respect to the cluster redshift. A mask is made to keep the program from fitting near the [SII], H$\alpha$ [NII], [OI], [NI], H$\beta$, and [OIII] emission lines. All spectra are masked near $\sim$6540\AA ~where an instrumental artifact appears, and for Abell~2457, there are residual sky lines from the sky subtraction process near 5600\AA, which are also masked from the fit.
As with all population synthesis codes, results vary with choice of IMF and base spectra and there is a well-known age-metallicity degeneracy. 

Thus, we do not use these results as an absolute measure of age and metallicity, but rather to investigate if there is a qualitative difference in spectra between the different regions in each of our targets.

\section{Results} \label{res}

\subsection{Age and Metallicity across the BCGs}

\begin{figure*}
\begin{minipage}[c]{0.32\textwidth}
\centering
\includegraphics[width=2.1in, height=1.5in, angle=0]{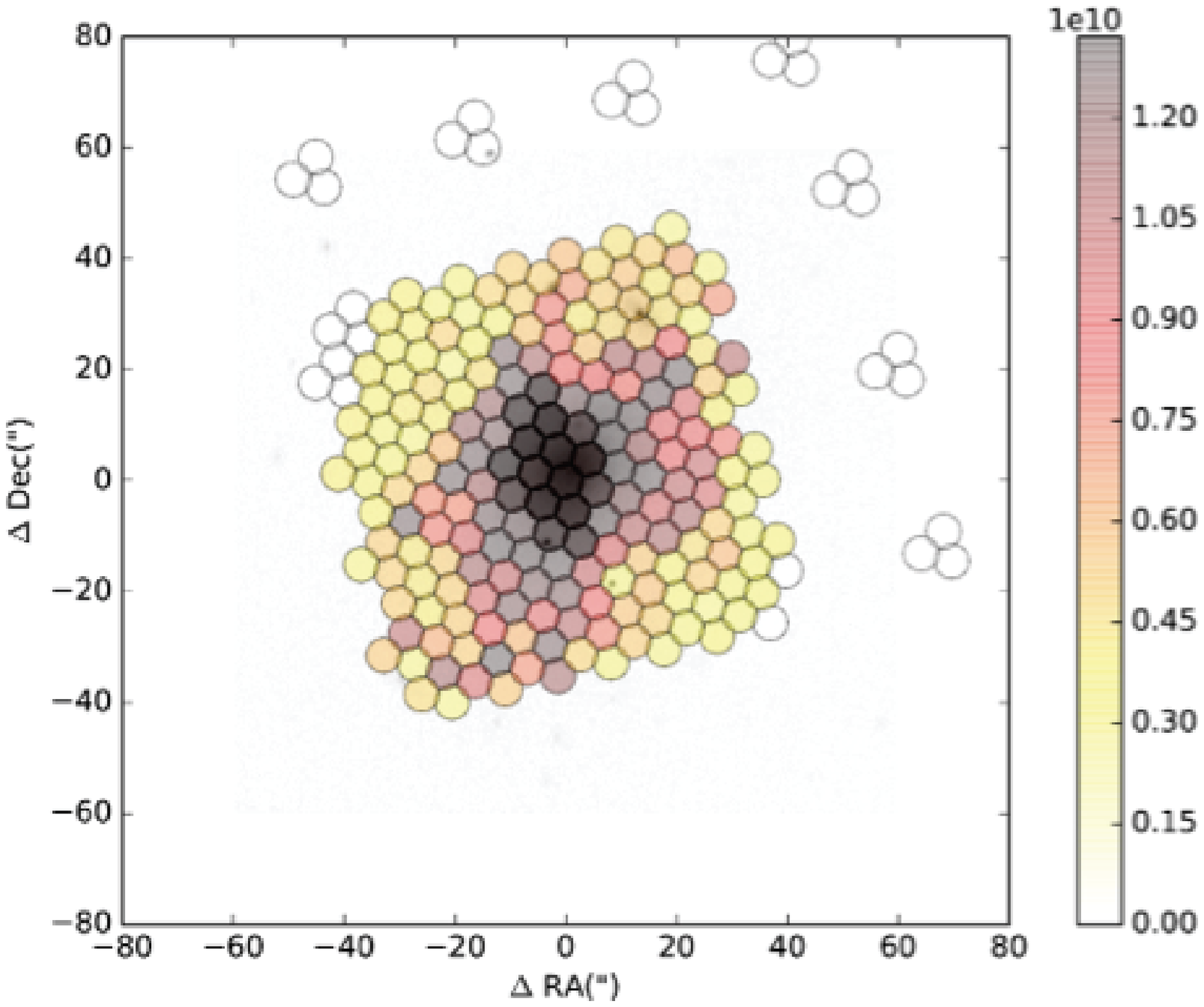}
\end{minipage}%
\begin{minipage}[c]{0.32\textwidth}
\centering
\includegraphics[width=2.1in, height=1.6in, angle=0]{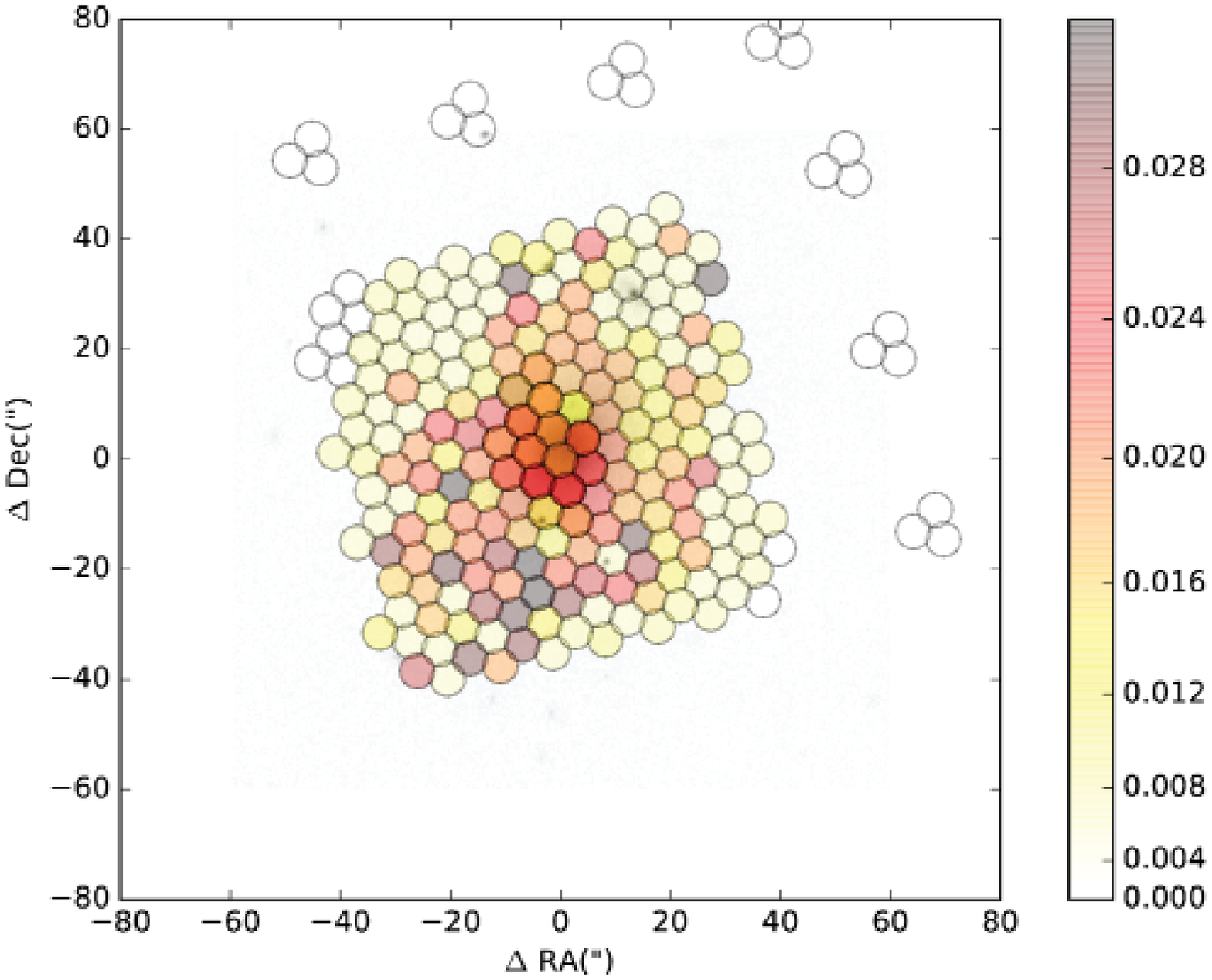}
\end{minipage}%
\begin{minipage}[c]{0.29\textwidth}
\centering
\includegraphics[width=2.1in, height=1.6in, angle=0]{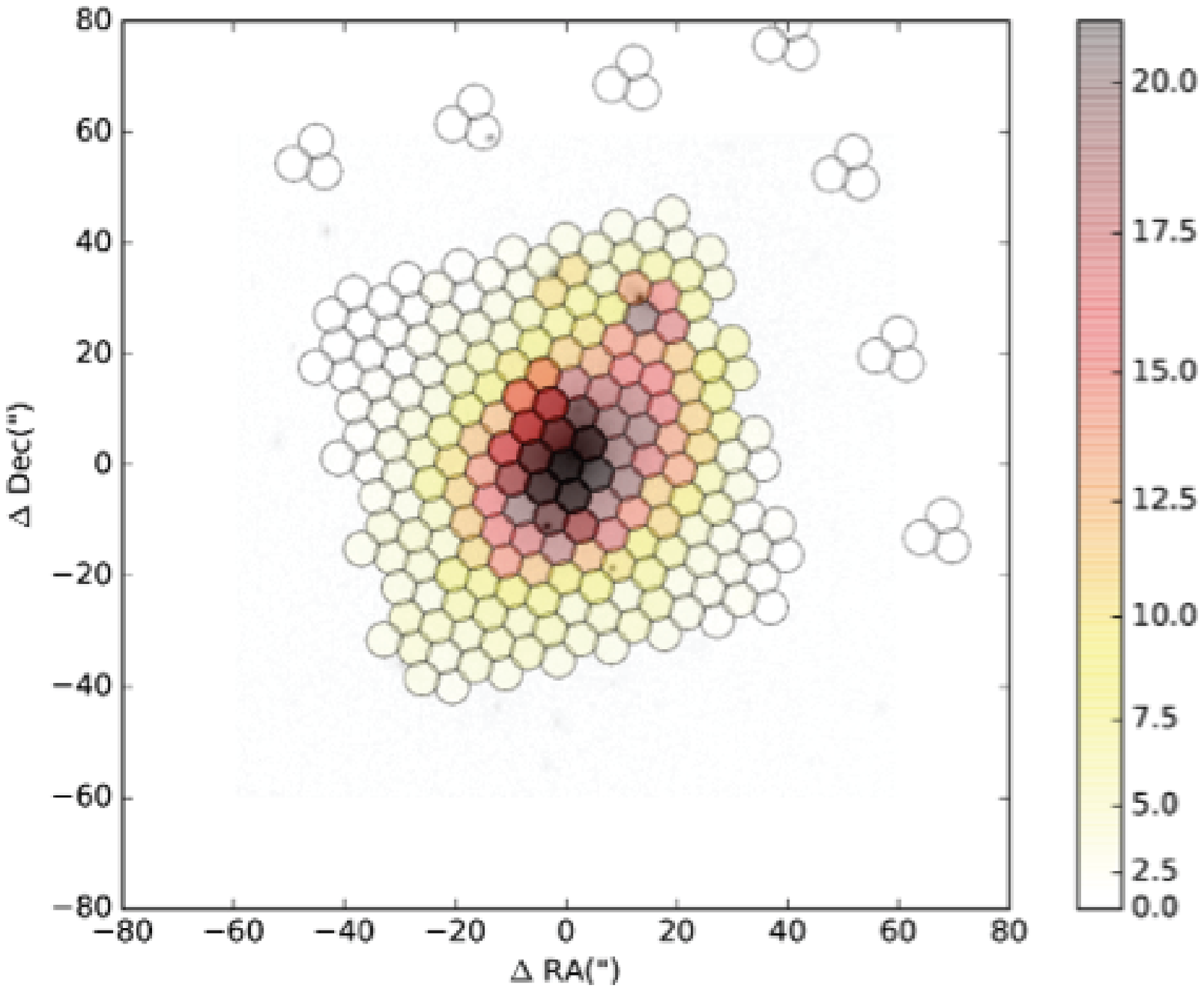}
\end{minipage}%
\\
\begin{minipage}[c]{0.32\textwidth}
\centering
\includegraphics[width=2.2in,height=1.6in,  angle=0]{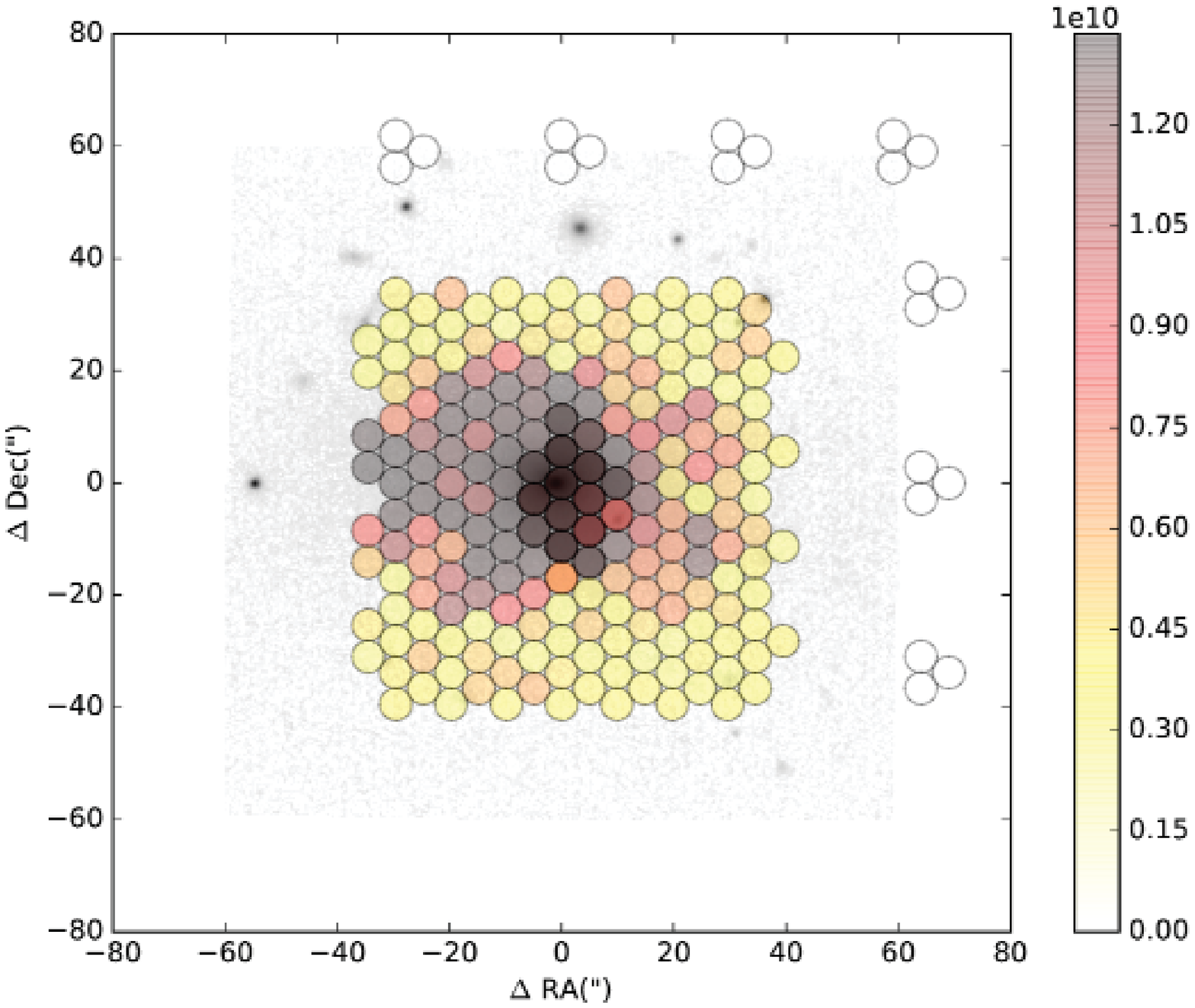}
\end{minipage}%
\begin{minipage}[c]{0.32\textwidth}
\centering
\includegraphics[width=2.2in,height=1.6in,  angle=0]{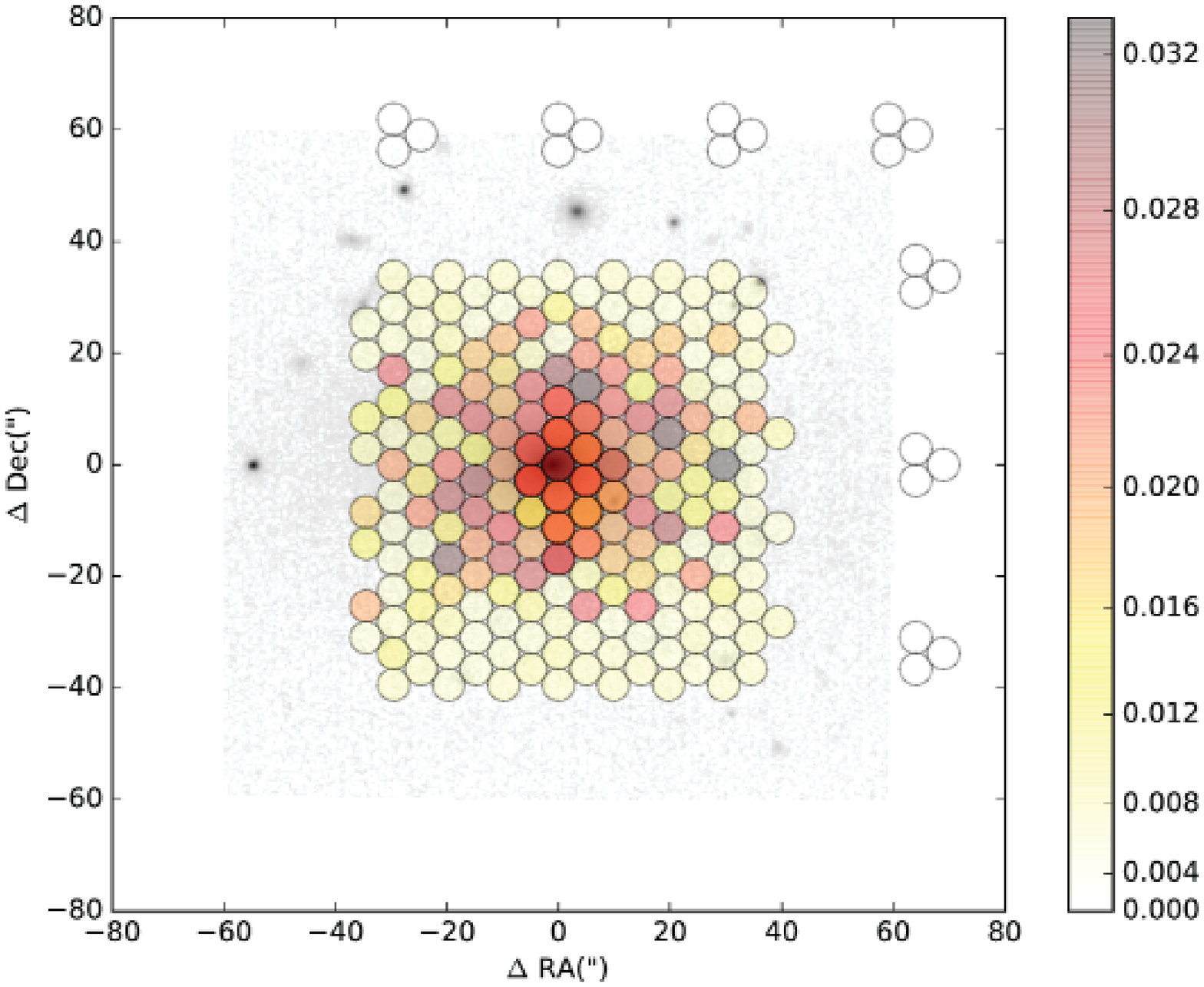}
\end{minipage}%
\begin{minipage}[c]{0.33\textwidth}
\centering
\includegraphics[width=2.2in,height=1.6in,  angle=0]{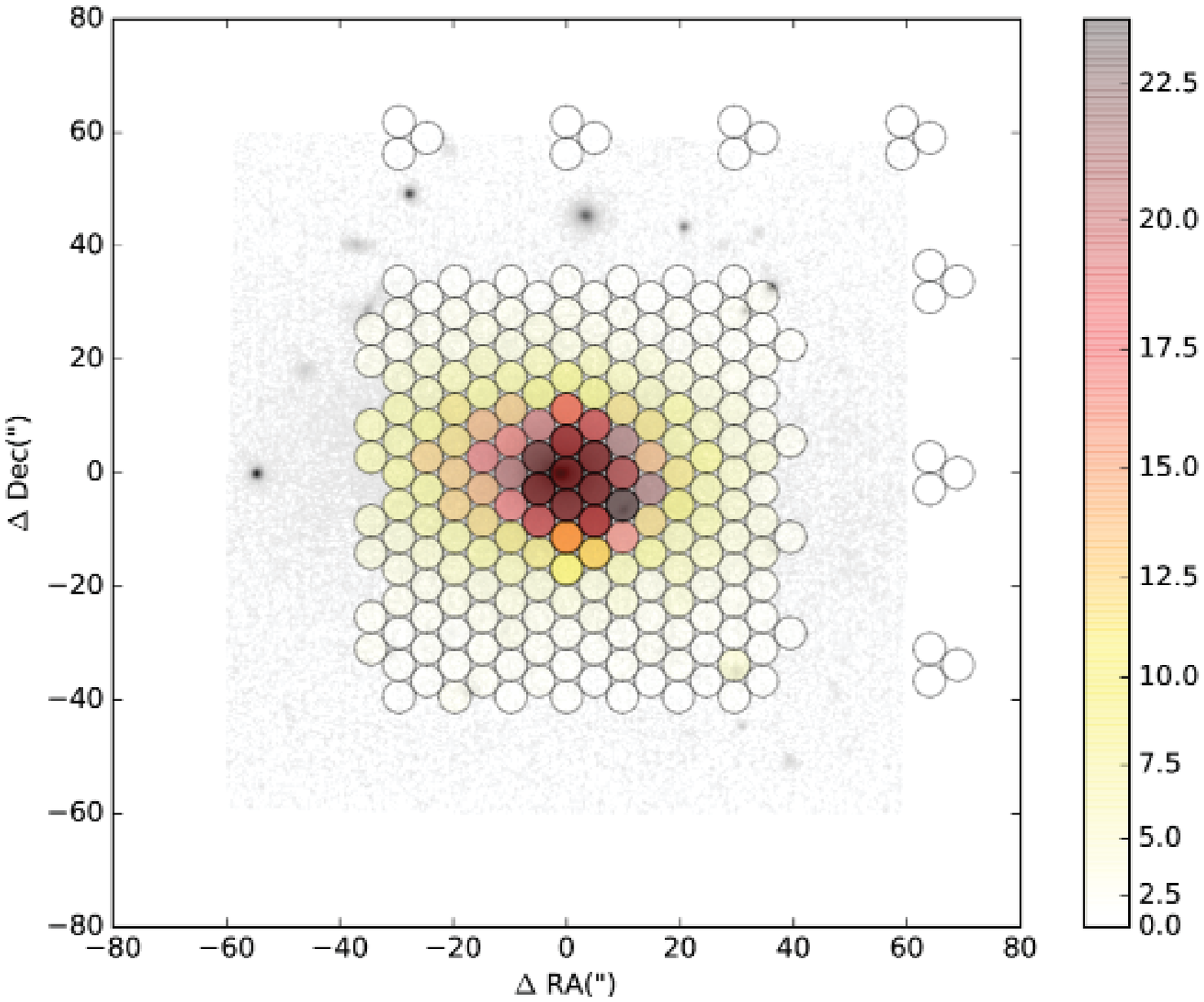}
\end{minipage}%
\\
\begin{minipage}[c]{0.32\textwidth}
\centering
\includegraphics[width=2.2in,height=1.6in,  angle=0]{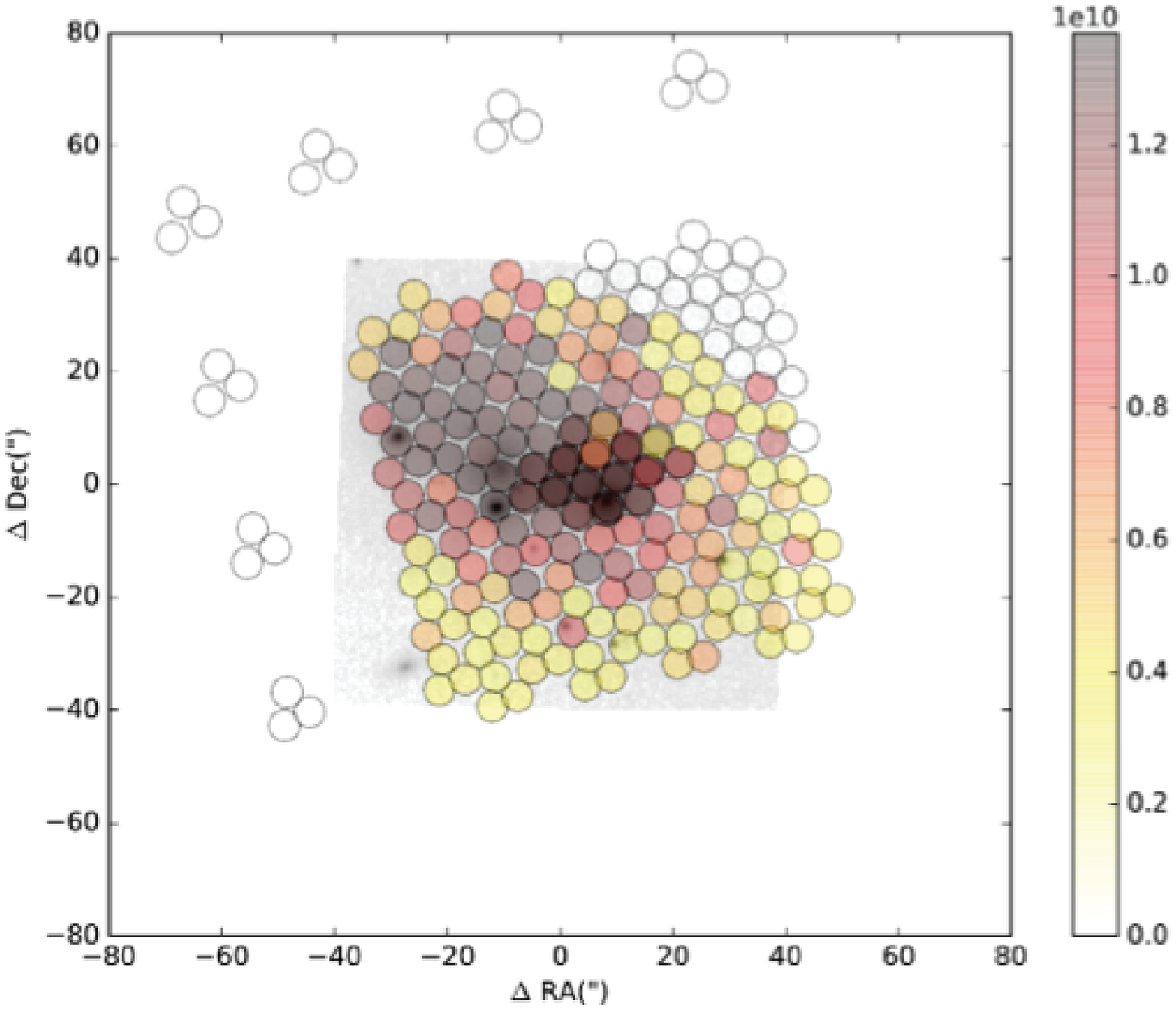}
\end{minipage}%
\begin{minipage}[c]{0.32\textwidth}
\centering
\includegraphics[width=2.2in,height=1.6in,  angle=0]{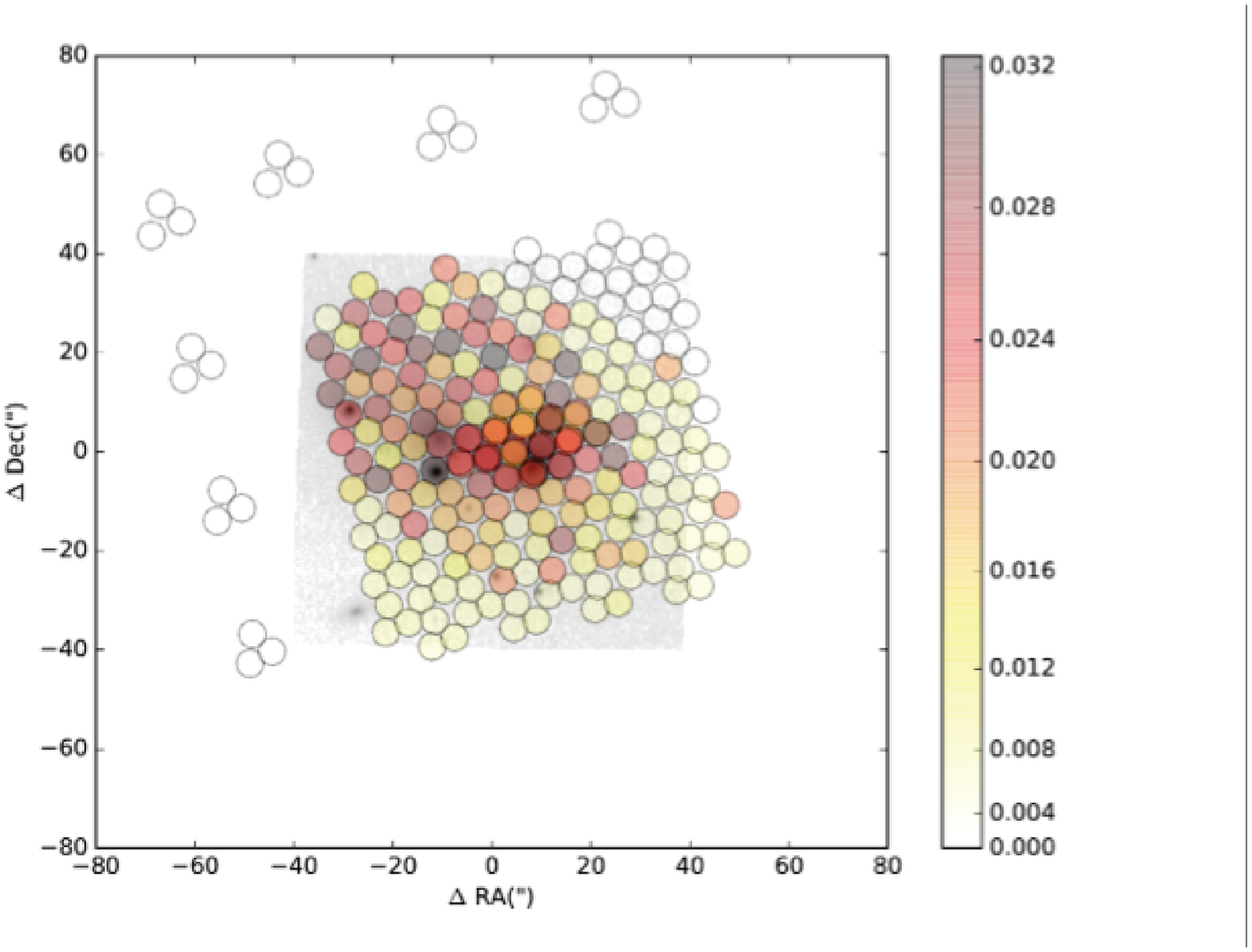}
\end{minipage}%
\begin{minipage}[c]{0.33\textwidth}
\centering
\includegraphics[width=2.2in,height=1.6in,  angle=0]{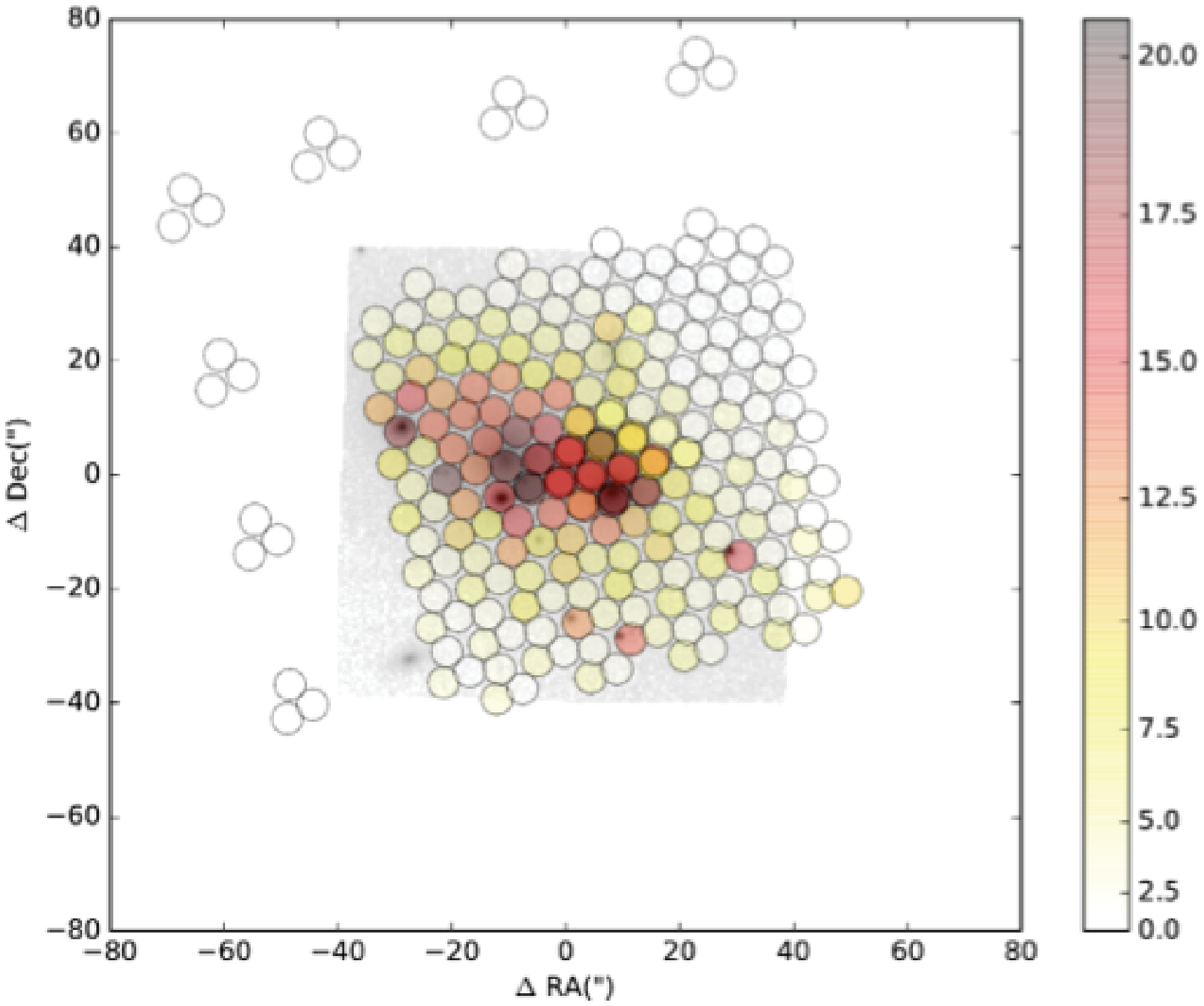}
\end{minipage}%
\caption {\textit{The best-fit age (left) metallicity (centre) and the signal to noise per fibre (right) are shown for Abell~85 (top), Abell~2457 (centre) and IIZw108 (bottom). The age and metallicity plotted are mass weighted averages of the four most abundant populations from the best-fit. North is up and East is to the left. Only the central fibres that receive light from the brightest cluster galaxies and companions have a signal to noise $>$7. Beyond, signal from several fibres must be combined in order to achieve a reliable spectrum. }}
\label{fig:Resagez}
\end{figure*}

A weighted average of the four most abundant populations as deterined by STARLIGHT is calculated for each fibre spectrum. The results are shown in Figure~\ref{fig:Resagez} which present the age (left), metallicity (centre), and signal to noise (right).

All fibres near the BCG are dominated by an old 13~Gyr stellar population. Near the core and BCG centre, fits include 60\% metal-rich stars (0.032) and 40\% metal-poor stars (0.006).  For Abell~85, the BCG core spectrum (Figure~\ref{fig:A85spectra}, left) shows that all of the bright H$\alpha$ emission is concentrated in the BCG core. It has a flux of 1.49$\pm$0.07$\times$10$^{-15}$erg/s/cm$^{2}$ and the high [NII]/H$\alpha$ ratio of 2.08$\pm$0.02 with [OIII]/H$\beta$ ratio of 0.33$\pm$0.02 is characteristic of a LINER spectrum \citep{bal81}.

Likewise, in Abell~2457 (Figure~\ref{fig:Resagez}, centre) the core (r$<$2.6$\,$kpc) and central (2.6$\,$kpc$<$r$<$7.8$\,$kpc) regions fit to models where all stars are old (age$>$13Gyr) and where the metallicity is dominated by a metal-rich population (82\% in the core, 58\% in the centre) but includes a population of old, metal-poor stars as well (0.006). The outer BCG populations include 30\% old, metal-rich stars and 63\% old, metal-poor stars. This lowers the weighted average for the metalicity in the outskirts (0.020).

For IIZw108 (Figure~\ref{fig:Resagez}, bottom), the BCG core, centre and outer regions also require 13Gyr stars. The metalicity is 60\% high (0.032), and 40\% low (0.006). Unlike other regions, where the best-fit is from the $\alpha$-enhanced stars, the outskirts of the IIZw108 BCG are fit best with 40\% $\alpha$/Fe=0.

\subsection{Stellar Populations in Companions and ICL}

\subsubsection{Identifying Regions for Combined Spectra}

\begin{figure*}
\begin{minipage}[c]{0.34\textwidth}
\centering
\includegraphics[width=2.6in, angle=0]{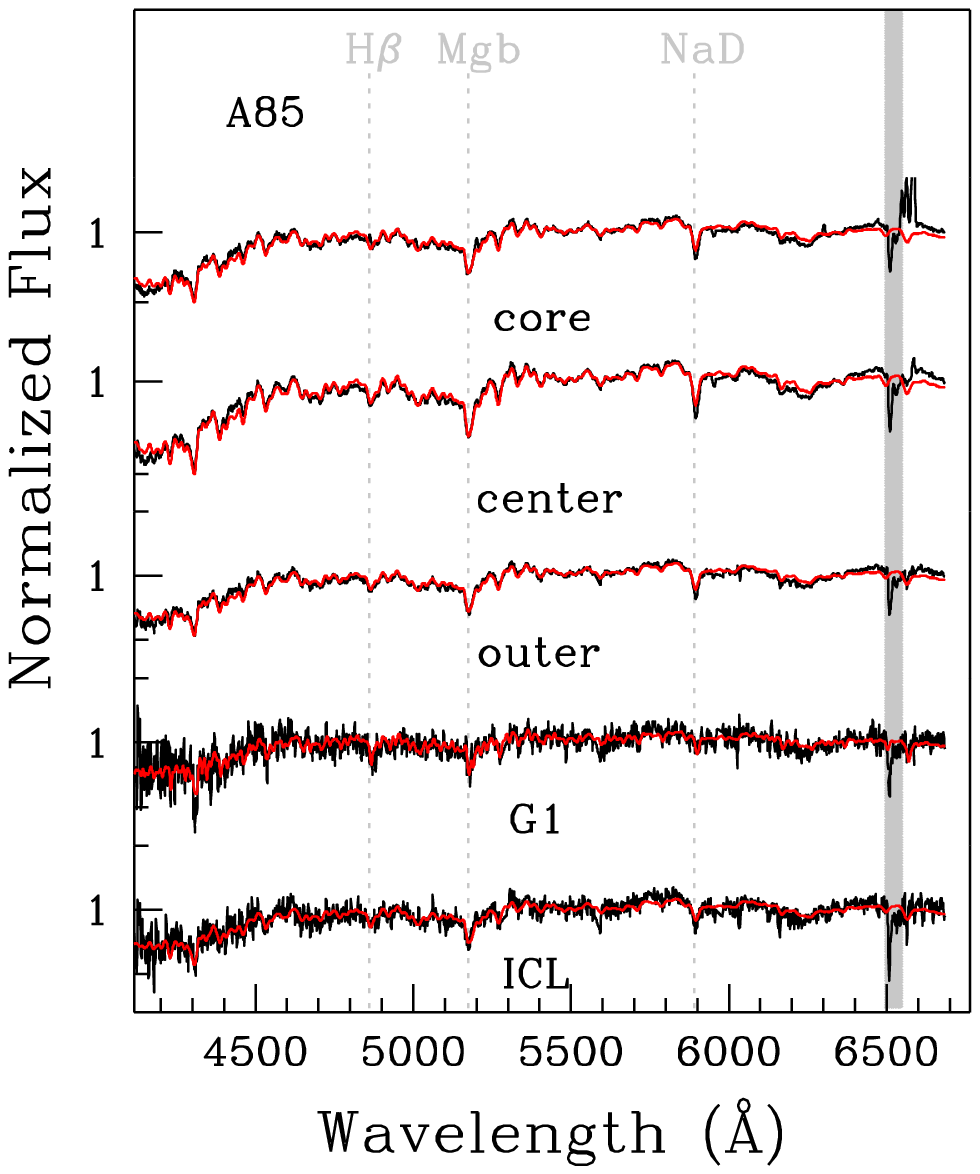}
\end{minipage}%
\begin{minipage}[c]{0.34\textwidth}
\centering
\includegraphics[width=2.6in, angle=0]{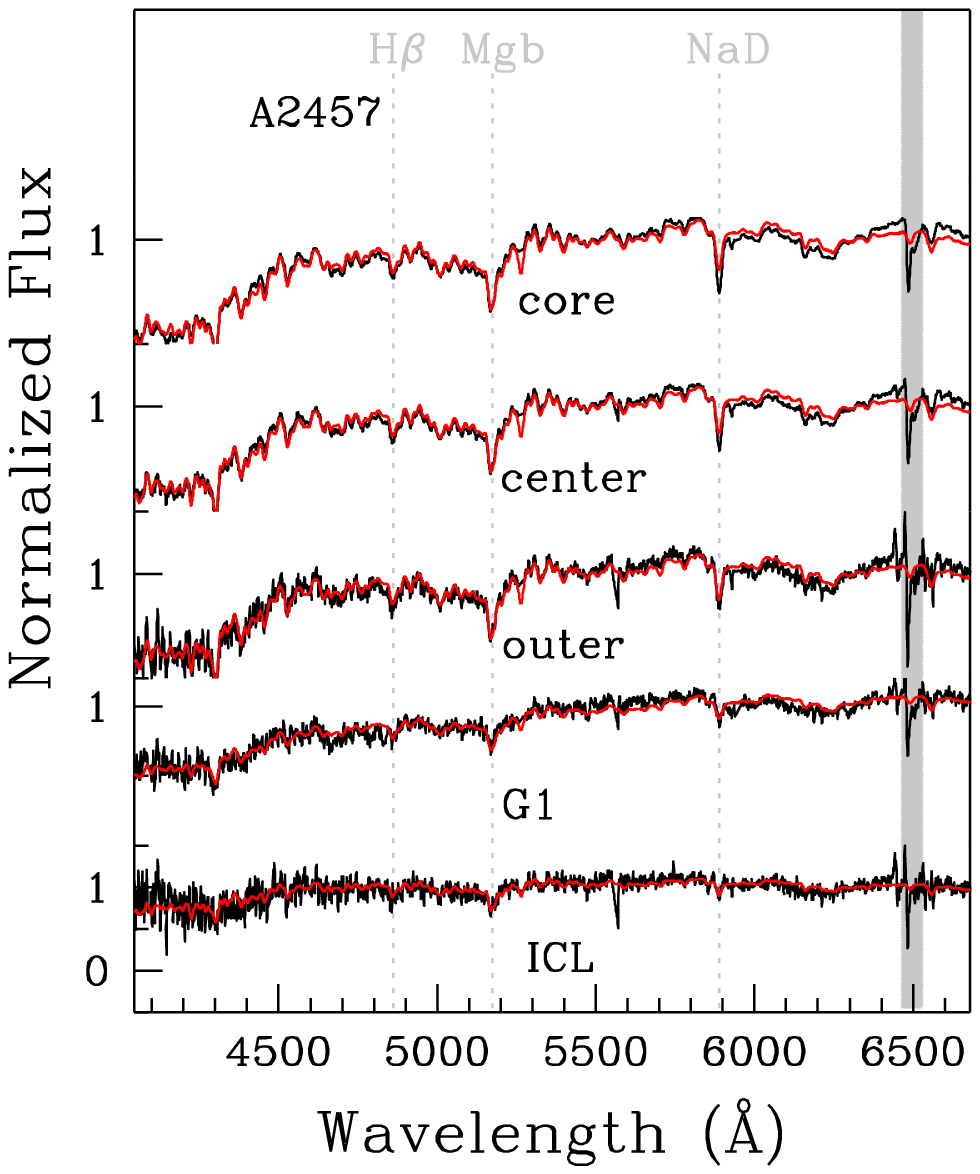}
\end{minipage}%
\begin{minipage}[c]{0.28\textwidth}
\centering
\includegraphics[width=2.6in, angle=0]{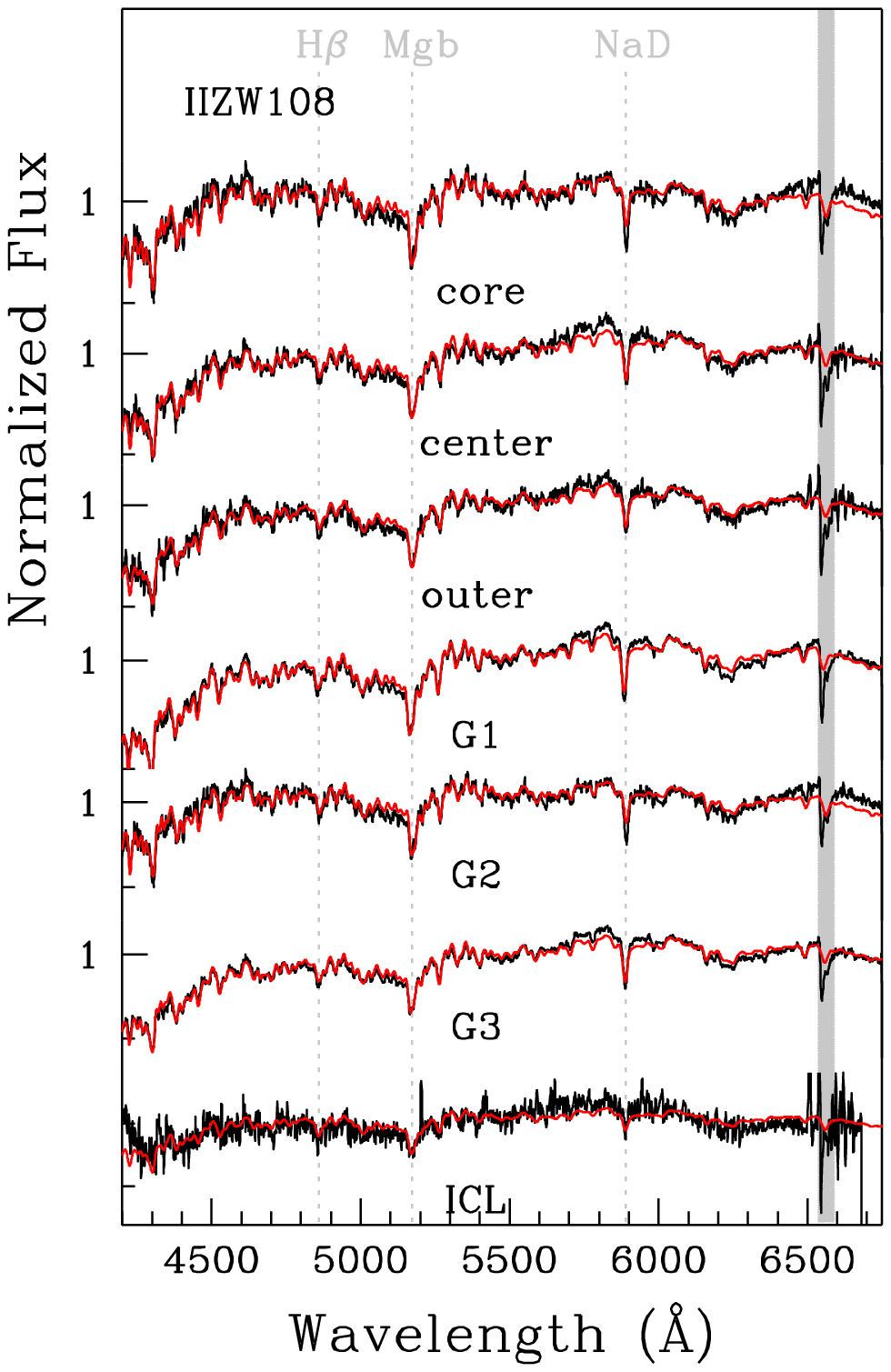}
\end{minipage}%
\caption {\textit{Combined spectra for cluster regions. The black line is the observed spectrum, and the red is the scaled best-fit model. In Abell~85 (left,) the BCG Core has very strong emission lines, indicative of a LINER. Neither Abell~2457 (centre) nor IIZw108 (right) show emission lines. After combining over 100 spectra, absorption lines in the ICL of all clusters are identified.}}
\label{fig:A85spectra}
\end{figure*}

To increase the spectral signal to noise to $\sim$50 in the BCG and $\sim$25 in the ICL, the r$^{\prime}$ images are used to define regions in which we median average the signal from multiple fibres. The BCG core includes light from our smallest spatial resolution element of one fibre diameter across. The BCG centre radially averages light one resolution element beyond that, and the BCG outskirts two units beyond this, as illustrated in Figure~\ref{fig:contmaps}. The companion galaxies are labeled with a G and are spectroscopically verified. Galaxies labeled with an I are interlopers. The median averaged spectra for these identified regions are shown in black in Figure~\ref{fig:A85spectra}. Population synthesis modelling as described in Section~\ref{sec:starlight} is performed and the best-fit model spectrum is overlain in red. The best-fit velocity, and velocity dispersion of the modelled spectrum is given in Table~\ref{tab:kins}.

Abell~85 - Figure~\ref{fig:contmaps} (left) shows one companion 41.3$\,$kpc north of the BCG (G1 in the top right panel) and one interloping galaxy to the northwest of the BCG (I1). The mean flux per fibre of the BCG core is 9.62 times greater than that of G1, so this is a potential minor merger. The SDSS photometric redshift has I1 at z$=$0.103$\pm$0.056, and we measure the galaxy velocity to be -1358$\,$km/s, placing it in front of the cluster core - though it could be an infalling galaxy on the edge of the cluster, seen in projection.  The populations for the companion G1 differ from the ICL and BCG, with populations most heavily weighted to young, metal-poor stars. 

Abell~2457 - Figure~\ref{fig:contmaps} (centre) shows one companion galaxy (G1) to the southwest of the BCG, a projected 15.3$\,$kpc from the BCG core. We find this galaxy has a velocity of -219$\,$km/s with respect to the cluster, within that of a typical rich cluster. G1 is the only region with a large fraction of young and metal-rich stars. It is reasonable to expect some contamination from the outer envelope of the BCG given the proximity of G1, indeed, the fit requires a population of old, metal-poor stars, also found in the outer regions of the BCG. 

IIZw108 - The right panel of Figure~\ref{fig:contmaps} reveals several bright sources seen in the r$^{\prime}$ images. With our spectra, we are able to identify three companion galaxies (G1-G3) in addition to the BCG. Two interloping galaxies are identified as well as several foreground stars. These are removed from the remainder of the analysis. All cluster galaxies are overwhelmingly dominated by an old (~13$\,$Gyr) stellar population. For the BCG and G2, the metallicity is split between a super solar and sub solar component. G1 and G3 and dominated by supersolar metallcities.  We choose the BCG to be the bright red galaxy that appears to be in the extended envelope of starlight following \citet{cra99}. We note that within the field of view of one fibre, the galaxies G1-G3 are all brighter than the BCG in r$^{\prime}$-band magnitudes and their light has a more concentrated profile.

\subsubsection{Defining the ICL}

We define the ICL as all fibres where there is a non-zero signal beyond the BCG outer region, excluding stars the companions. The results in at least 100 fibres for each clusters and allows for a singal to noise of the combined spectrum of $\sim$25.

In recent simulations, the ICL has been defined as following the cluster potential rather than that of the BCG. But, with the exception of only one system \citep{mel12} all previous observatinal studies of the ICL have been conducted using photometric data. The  majority register ICL as faint light that extends up to 100s of kpc from the cluster core \citep{kri06,mih05,mon14}. This discrepancy makes it difficult to characterize exactly the ICL. It has been shown \citep{pre14} that the traditional method of identifying ICL through arbitrary surface brightness limit cuts \citep{kri06}, are subject to contamination from the outskirts of large cluster galaxies. It is especially difficult to separate the BCG from the core of the ICL, so many authors consider the two together. \citet{pre14} have found that fitting deep photometric data shows breaks in the light profile at $\sim$40$\,$kpc and also find that the ICL contribution is more imporant close in to the BCG than further out beyond $\sim$150$\,$kpc. This is also roughly the region that \citet{mel12} and \citet{coc10} use to define the ICL for their spectroscopic measurement. We effectively include light from $\sim$25-40$\,$kpc, so are likely contaminated by the BCG outskirts. It is possible that further work such as analysis of surface brightness profiles from deep imaging of the cluster core may prove our ICL region to be the furthest reaches of BCG outskirts. That being said, the fact that the ICL stellar populations change so drasitcally (discussed below) supports the notion that we are indeed identifiying a different population of stars.

\subsubsection{Contrasting ICL and BCG Stellar Populations}

\begin{figure}
\centering
\includegraphics[width=3.6in, angle=0]{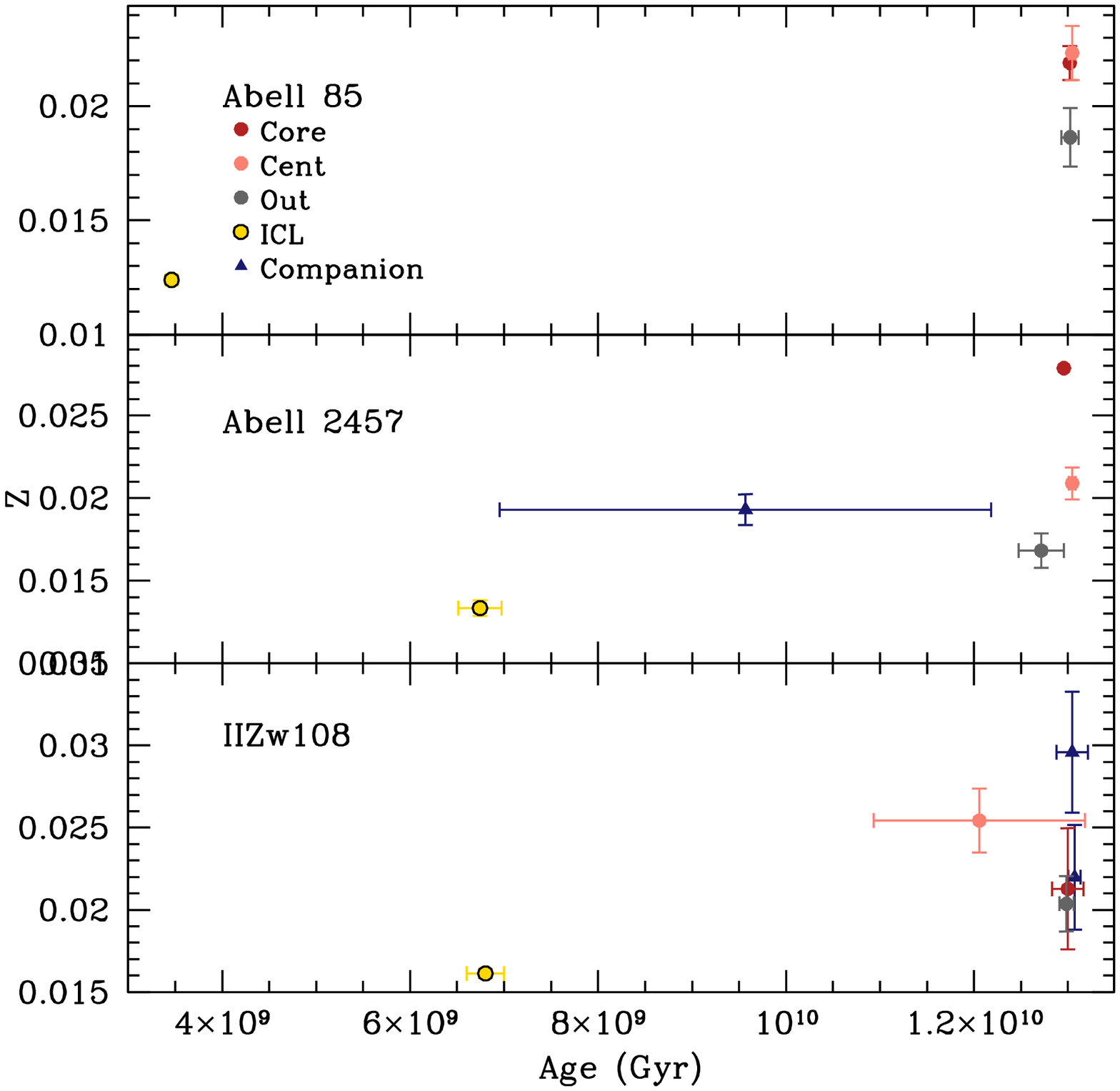}
\caption {\textit{Spectral synthesis results of age and metallicity. The average age and metallicity for each region are plotted for IIZw108 (top), Abell~2457 (centre), and Abell~85 (bottom). The BCG core and centre are generally more metal-rich and older than the outskirts, and the ICL is always the youngest, most metal-poor population. Errorbars are the standard deviation from different runs with slight changes in the input spectra. Errors for the Abell~2457 companion are particularly large as the spectrum is made of only 2 fibres.}}
\label{fig:Modagez}
\end{figure}

The ICL, in all three of our galaxies, shows much younger ages and lower metallicities than any region of the BCG and the core and centre of the BCG are typically home to the oldest, most metal-rich population. This can be seen in  Figure~\ref{fig:Modagez} where we plot the average stellar populations from the best-fit synthesis model for each region in the three clusters. The errorbars are calculated by finding the standard deviation of solutions obtained from 5 sets of spectral fits - each input spectrum differing by randomly removing 5-10\% of the fibres from the median average and re-running the model fitting. The exception is the companion galaxies which comprise of only a few fibers, in which case only 3 fits were run, and the errors are larger. The values of age and metallicity vary quite a bit when using different models, but we found the overall trends to be consistant when the spectra were fitted with \cite[Coelho et al. 2007]{coe07} and \cite[Bruzual \& Charlot 2003]{bru03} models. 

We can learn more by examining the relative fraction of the most abundant populations in each cluster. In all three cases, the ICL is fit with a large population of $\alpha$-enhanced ($\alpha$/Fe = 0.40) old, metal-rich stars like the BCG, but also includes a significant population of young, metal-poor stars, in stark contrast to the BCG. This is illustrated in  Figure~\ref{fig:Modagez} where the average stellar populations from the best-fit synthesis model for each region in our three clusters are plotted. 

In Abell~85 the ICL is best-fit with 35\% young (5$\,$Gyrs,)  metal-rich (Z=0.032) stars, 34\% old (13$\,$Gyrs,) metal-poor (Z=0.006) stars and 31\% young (5$\,$Gyrs,) metal-poor stars (Z=0.006).

In the case of Abell~2457, the ICL is best-fit with 47\% young (5$\,$Gyrs,)  metal-poor (Z=0.006) stars, 37\% old (12$\,$Gyrs,) metal-rich (Z=0.032) stars and 12\% old (12$\,$Gyrs,) metal-poor stars (Z=0.006).

In IIZw108, the ICL is the only region to show any significant young population. It is best-fit with 55\% young (3$\,$Gyrs,)  metal-poor (Z=0.006) stars, 48\% old (12$\,$Gyrs,) metal-rich (Z=0.032) stars.

\begin{table}
\centering
\caption{Best-fit model population results. The velocity of the region with respect to the cluster redshift is in column 2, and with respect to the BCG is in column 3. Columns 4 and 5 show the measured velocity dispersion and that corrected for instrumental broadening, respectively. Column 6 lists the dynamical mass, as calculated in the text.}
\begin{tabular}{l c c c c l}
\hline \hline
Region & $\Delta$$v_{Cl}$ & $\Delta$$v_{core}$ & $\sigma$$_{meas}$& $\sigma$$_{cor}$  &M$_{dyn}$\\
& km/s & km/s & km/s & km/s  &$ \times$10$^{11}$\\
& & & &  &M\sun\\
\hline 
Abell 85&&&&\\
BCGcore & 97 & 0 & 369 & 365 &3.81\\
G1 & 373 & 276 & 181 & 172 & 0.85\\
ICL & 150 & 53 & 351 & 346 &\\

Abell 2457&&&&\\
BCGcore & -230 & 0 & 375 & 371 &3.94\\
G1 & -219 & 11 & 398 & 394 & 4.45\\
ICL & -178 & 52 & 279 & 274&  \\

IIZw108&&&&&\\
BCGcore & -38 & 0 & 288 & 283 &  2.29\\
G1 & -432 & -394 & 334 & 330  &3.11\\
G2 & -38 & 0 & 287 & 282 &2.27\\
G3 & -226 & -188 & 328 & 324 &3.00\\
ICL & -180 & -142 & 389 & 385 &  \\
\\
\hline
\label{tab:kins}
\end{tabular}
\end{table}

\subsection{Companions Kinematics: Dynamical Friction Timescale}

The best-fit model kinematics are shown in Table~\ref{tab:kins}. The velocity offsets between the best-fit and observed region spectra verify spectroscopic companions and are used to calculate the dynamical friction timescale, t$_{fric}$ \citep{bin87}, between any close companions and the BCG.

\begin{equation}
t_{fric} = \frac{2.64\times 10^5 v_{vir}r^2_{sat}} {m_{sat}} \Lambda
\label{dynf}
\end{equation}

Where, $v_{vir}$ is the velocity offset with respect to the BCG core velocity. The error on the velocity and velocity dispersion has been estimated by fitting the observed spectra to many runs, varying base spectra, and numbers of base spectra. The velocity error is on the order of $\pm$10$\,$km/s and for the velocity dispersion it is $\pm$20$\,$km/s. The separation between the galaxies, $r_{sat}$, is measured from the SDSS images, and as it is in projection, results in a minimal timescale. To estimate the satellite galaxy mass, $m_{sat}$, we calculate the dynamical mass from the corrected velocity dispersion using the virial theorem. The effective size of local BCGs has been measured to be 4.5$\pm$1.0$\,$kpc \citep{von06}, which, at these local redshifts, matches well with the fibre size of 4.67$^{\prime\prime}$. 

For Abell~85, the relative mass and rather large separation of the companion results in a long dynamical friction timescale of 2.5$\,$Gyr, and the systems are unlikely to merge soon. The small difference in velocity between G1 and BCG in the case of Abell~2457 allows for a short dynamical friction timescale (1.5$\,$Myr). All three of the companions for IIZw108 are within 20$\,$kpc of the BCG, and likewise, all have short dynamical friction timescales calculated to be less than 0.1$\,$Gyr.

Table~\ref{tab:kins} shows that the BCG core and ICL are offset in velocity, most significantly in IIZw108. This could be from residual motion in the ICL light from recent mergers. Although, the difference between G3, which has the a larger dynamical mass, and the ICL is much less. 

\subsubsection{Companion Kinematics: Are the companions bound?}

We calculate the probability of the companions being bound to the BCG. Following \citet[]{bee82,bro06a,jim13} we require:

\begin{equation}
V_{r}^{2} R_{p} \leq 2GM_{dyn}sin^2\alpha cos\alpha 
\label{bound}
\end{equation}

Where, $V_{r}$ is the relative velocity between companion and the BCG from Table~\ref{tab:kins}, $R_{p}$ is the projected distance between the two measured on the SDSS r$'$ images, $\alpha$ is the angle between the galaxies, and M$_{dyn}$ is dynamical mass as defined below \citep{cap06}:

\begin{equation}
M_{dyn} = 5R_{e}\sigma^{2}/G 
\label{bound}
\end{equation}

Where R$_{e}$ is the effective radius, measured on the sloan r$'$ image using SExtractor \citep{ber96}, $\sigma$ is the corrected velocity dispersion of the galaxy from Table~\ref{tab:kins}, and G is the gravitational constant.

We find that for Abell~85, G1 is likely to be bound only at the 25\% level, for Abell~2457 G1 is likely to be bound at the 93\% level, and for IIZw108 the companions are likely to be bound at the 20, 95, and 54\% level, respectively.

\section{Discussion}

In this section, we compare our results of stellar populations and kinematics to those of similar studies in the literature in order to verify if our results are commensurate with the galaxy growth predictions of the current $\Lambda$CDM paradigm.

\subsection{BCG stellar populations}\label{spops}

Our measurements have shown that the cores of the BCGs in all our clusters are dominated by old, metal-rich populations, consistent with predictions from \citep{ton12} as well as observations from \citet{lou09} and \citet{von06}, who use single stellar population fits. We do find a fraction of metal-poor stars, similar to \citet{bro07}, who find a range of metallicities. 

In line with \citet{oli15}, we do not measure any strong metallicity gradients in the cores of the BCG.  We interpret this in the same way as \citet{bro07}, who found no universal trend in metallicity gradients within their sample of 6 BCGs using GMOS long slit spectrograph and fitting single stellar populations - though generally, the metallicity and age are highest at the core of their BCGs and occupy a narrow range of old stars (11-15$\,$Gyr). Their gradients were measured out to 6.5-30$^{\prime\prime}$ (our `BCG centre'-`BCG outer'). We concur with \citet{bro07} that this likely reflects differences in individual system's assembly history. However, both Abell~85 and Abell~2457 - the two clusters with clear BCGs - show outer regions that are fit with a higher fraction of low metallicity stars than the inner regions.  At this point, the BCG and ICL are both growing by a component from satellite galaxies \citep{fit15}, or from smaller, younger, more metal-poor galaxies \citep[]{con14,dem15}. 

The above is consistent with the picture of \citet{del07} and \citet{tof14}, where the stars that make up the BCG cores formed long ago \citep{guo08} with subsequent rapid growth \citep{dub98,lap13}, by dry mergers \citep{mil05,blu12,ske12}, and, in the absence of strong cool cores, with little new star formation seen. This is also consistent with what is expected from downsizing, where the most massive galaxies formed first and thus have an older stellar population with a longer timescale in which to enrich the interstellar medium. Over time, smaller galaxies, made of younger stars merge and add a mix of populations to the outer BCG envelope and the ICL, like we see in Abell~85. The fact that Abell~2457 has substructure present in the X-ray suggests recent activity, but it was long enough ago that the optical light had time to smooth itself over. Though small (3\%), a young, metal-poor population is seen in G1, lending evidence to the idea that minor mergers with cluster galaxies can add stars to the ICL. 

For IIZw108, only the ICL shows a component of younger stellar populations. At this stage of imminent major-mergering, the ICL hosts properties completely unidentified in the massive galaxies. It must have been formed by a population that is not shared with the merging galaxies that will form the final BCG core. Over time, some of this young population will age and could be the source of the old, metal-poor populations seen in Abell~85 and Abell~2457.

\subsection{Merging and Close Companions}

IIZw108 shows strong evidence for recent and ongoing major mergers, as four central galaxies could vie for the title of BCG given their large masses and red colours. Using the central galaxy as the BCG, G1 has a low probability of being bound, G2 and G3 are likely bound, and the merger timescales are all of the order of $<$0.1$\,$Gyrs. The ICL has a higher velocity dispersion than the BCG and Galaxies G1-G3, evidence that the cluster potential is domiant and that the ICL is unrelated to the major merger. \citet{tre77,loh06} and \citet{smi10} have all found that if the luminosity gaps between the 1st and second rank galaxy are low, it is indicative of a recent merger. Indeed, this is the case with IIZw108 where the brightest galaxies have $\Delta$m$_{r} = $0.67 (SDSS). This system is similar to the case of Abell 407 (z=0.046), another local system with several massive red galaxies at the centre \citep{sch82}, which appears to be undergoing BCG formation \citep{mac92,edwards15}. These two systems illustrate what may be common at high redshift, and show that build-up of BCGs by dry major mergers still occurs locally. Moreover, \citet{bro11} also analysed a sample of BCGs and their companions, finding evidence that major merging continues in the local universe. Semi-analytic models of \citet{ton12} suggest that BCG mass growth continues after z$<$1, and indeed, cosmological models must be able to form mostly large cD galaxies locally, while still allowing for cases like these two where massive large-scale 1:1 mergers are still happening.

Most of the major mergers in the Abell~2457 system are likely far in the past, but the short dynamical friction timescale of G1, and that fact that it is likely bound, implies a minor merger could be in process.   Abell~85 shows no evidence for current growth or assembly of the BCG, as neither cluster core shows evidence for recent disruption by merging and given that the velocities of the BCG core and outskirts and ICL are all similar. Furthermore, Abell~85 shows no evidence for the potential of imminent minor mergers as G1 in this system has a long merger timescale and is unlikely to be bound.

\subsection{Intracluster Light} \label{iclsec}

We have extracted the ICL components from our three galaxy cluster cores and fit model stellar populations to the spectra. This exercise has found that the ICL spectra are better fit with a young, metal-poor component, not seen in the BCG and large red galaxies.  This supports a different origin for the inner BCG and ICL. 

We compare our results to the spectroscopic observations of the BCG and ICL for RXJ0054.0-2823 \citep{mel12,tol11}, where the ICL component was measured by combining several observations at $\sim$50$\,$kpc from the BCG core. It was found to have the largest contribution from a population of old and metal-rich stars, with a 16\% contribution from old, metal-poor stars. A young, metal-poor component could comprise at most $\sim$1\%. Although this galaxy is at a higher redshift (z$\sim$0.29) than our local systems, the results of \citet{con14} show that $>$$\,$60\% of the ICL is in place by z=0.3, and the GAMA results suggest little BCG growth has taken place since z=0.27. The RXJ0054.0-2823 result is similar to IIZw108 and Abell~2457, in that the ICL is fit with a component of old, metal-rich stars like in the BCG. However, we also find that a better fit includes a population of 30-50\% young, metal-poor stars. The difference between our local clusters could be explained if the metal-poor component is being recently added by lower-mass galaxies, though we hesitate to generalize from comparing to just one source.

The models of \citet{puc10} predict that z$\sim$0 clusters should have an ICL with a 30\% young stellar component. This agrees with our data.

The ICL also requires a population of stars that mirror those of the BCGs. It has been suggested \citep{con07,bur12,oli14,zha16} that if much of the stellar mass in BCG mergers is also shared with building the ICL, this could explain why observations of BCGs today show less mass than expected from simulations. That we see shared stellar populations in the BCG and ICL supports this hypothesis. 

\section{Conclusions} \label{con}

In this paper, we have reduced and analysed the data of a 75$\times$75$\,$kpc$^{2}$ region around three BCGs that are part of a larger set of 16 galaxy cluster cores. We measure the stellar populations and kinematics for the BCG core, outskirts, and close companions. Our survey design allows for the first ever measurements of stellar populations of the ICL using IFU data. With instruments such as SITELLE and MUSE now online, this could prove to be an efficient method.

The dominant populations throughout are of old, metal-rich stars, and in particular for each of the three clusters, it is always the most massive galaxy that has the highest fraction of old, metal-rich stars, and the ICL requires a more varied mix that includes younger and metal-poor stars.  Abell~85 shows no close companions that are bound and likely to merge. Abell~2457 finds evidence for one potential minor merger. IIZw108 is found to have three bright galaxies in addition to the BCG, and two are likely bound and could merge within $<$0.1$\,$Gyr. 

The metallicity and ages discussed above and properties of close companions are consistent with the three BCGs being in different phases of formation. Abell~85 is smooth in the optical and relaxed in the X-ray. It also has the largest difference in stellar populations between the BCG core and outskirts and the ICL. There has been more time between when the BCG built most of its mass and now for infalling galaxies far from the core to collect into the ICL. Abell~2457 is also smooth in the optical, but there exist relics in the X-ray wavelengths. Such relics suggest that there must have been recent group-scale mergers in the past that occurred long enough ago for the optical light to appear smooth today. Abell~2457 shows a smoother change in stellar populations going from the BCG core, through the outskirts and to the ICL. The low-mass close companion has the ages and metallicites between that of the BCG regions and the ICL. At this stage, the minor mergers building the BCG are contributing to the ICL. Finally, IIZw108 shows little extended ICL, with a population of young, metal-poor stars that do not exist in any of the other merging galaxies. This system is currently undergoing major merging and is in the process of building up its BCG. These major mergers do not contribute to the young, metal-poor component of the ICL, but part of this population will age over time and may be the source of the old, metal-poor component in systems like Abell~85 and Abell~2457.

\section*{Acknowledgments}
This study uses data gathered with SparsePak on the WIYN Telescope at Kitt Peak National Observatory. We would like to thank instrument scientist Jenny Power and telescope operators Krissy Reetz and Dave Summers for their valuable assistance in obtaining observations. We acknowledge support from the STARS program at Yale and the CT Space Grant Consortium as well as Anthony Festa (and the Yale Summer Science Research Institute program) and Vasilije Dobrosavljevic for their contributions to the project. We also wholeheartedly thank Eric Wilcots and Sean McGee for fruitful discussions, and the referee for thoughtful comments that lead to improvements in the paper. This research has made use of the NASA/IPAC Extragalactic Database (NED), which is operated by the Jet Propulsion Laboratory, California Institute of Technology, under contract with the National Aeronautics and Space Administration. Funding for the SDSS DR10 has been provided by the Alfred P. Sloan Foundation, the Participating Institutions, the National Science Foundation, the US Department of Energy, the Japanese Monbukagakusho, the Max Planck Society. The SDSS is managed by the Astrophysical Research Consortium for the Participating Institutions (see list at http://skyserver.sdss.org/dr10/en/credits/). The STARLIGHT project is supported by the Brazilian agencies CNPq, CAPES and FAPESP and by the France-Brazil CAPES/cofecub program.
\bibliographystyle{mnras}
\bibliography{le1}

\bsp	
\label{lastpage}
\end{document}